

\input mn
\input epsf


\let\sec=\section
\let\ssec=\subsection


\def\bigstrut{\vrule width0pt height0.35truecm}
\font\japit = cmti10 at 11truept
\def\ss{\scriptscriptstyle\rm}
\def\ref{\parskip =0pt\par\noindent\hangindent\parindent
    \parskip =2ex plus .5ex minus .1ex}
\def\gs{\mathrel{\lower0.6ex\hbox{$\buildrel {\textstyle >}
 \over {\scriptstyle \sim}$}}}
\def\ls{\mathrel{\lower0.6ex\hbox{$\buildrel {\textstyle <}
 \over {\scriptstyle \sim}$}}}
\newcount\equationo
\equationo = 0

\newcount\fred
\fred=0

\def\outeqn#1{\the #1}

\def\leftdisplay#1$${\leftline{$\displaystyle{#1}$
  \global\advance\equationo by1\hfill (\the\equationo )}$$}
\everydisplay{\leftdisplay}

\def\kmsmpc{\;{\rm km\,s^{-1}\,Mpc^{-1}}}
\def\hompc{\,h\,{\rm Mpc}^{-1}}
\def\mpcoh{\,h^{-1}\,{\rm Mpc}}

\def\ig#1#2#3{
\beginfigure{#1}
\epsfxsize=8.4cm
\ifnum #2 = 1 \epsfbox[28 186 488 590]{pfig#1.ps} \fi
\ifnum #2 = 2 \epsfbox[53 15 465 785]{pfig#1.ps} \fi
\ifnum #2 = 3 \epsfbox[101 15 434 790]{pfig#1.ps} \fi
\caption{%
{\bf Figure #1.}
#3
}
\endfigure
}

 

\def\annrev{ARA\&A}

\def\apj{ApJ}
\def\apjs{ApJS}
\def\mn{MNRAS}
\def\nat{Nat}

%

\pageoffset{-0.8cm}{0.2cm}




\begintopmatter  

\vglue-2.2truecm
\centerline{\japit Accepted for publication in Monthly Notices of the R.A.S.}
\vglue 1.7truecm

\title{The evolution of galaxy clustering}

\author{J.A. Peacock}

\affiliation{Royal Observatory, \bigstrut Blackford Hill, Edinburgh EH9 3HJ}

\shortauthor{J.A. Peacock}

\shorttitle{The evolution of galaxy clustering}


\abstract{%
This paper investigates whether nonlinear gravitational instability
can account for the clustering of galaxies on large and small scales, and for
the evolution of clustering with epoch.
The local clustering spectrum is accurately established,
and bias is not a great source of uncertainty:
the real-space power spectra of optical and IRAS
galaxies show only a weak
scale-dependent relative bias of $b \simeq 1.15$ on large
scales, increasing to $b \simeq 1.5$ on the smallest scales.
Comparison with results in redshift space favours a
relatively small distortion parameter
$\beta_{\rm opt}\equiv \Omega^{0.6}/b_{\rm opt} \simeq 0.4$.

No CDM-like spectrum is consistent with the shape of the
observed nonlinear spectrum. Unbiased low-density models
greatly overpredict the small-scale correlations;
high-density models would require a bias which does
not vary monotonically with scale. The true linear
power spectrum contains a primordial feature at $k\simeq 0.1 \hompc$,
and must break quite abruptly to an effective
slope of $n\ls -2.3$ on smaller scales

This empirical fluctuation spectrum also fits
the CFRS data on the evolution of clustering,
provided the universe is open with $\Omega\simeq 0.3$.
Only this case explains naturally how the small-scale spectrum can
evolve at the observed rate while retaining the same
power-law index. An unbiased open model also
matches correctly the large-scale COBE data, and offers
an attractively simple picture for the phenomenology
of galaxy clustering.
}


\maketitle  

\sec{INTRODUCTION}

Studies of galaxy clustering have progressed steadily
towards measurements on ever larger scales.
The influential text by Peebles (1980) could call only on
results from angular clustering in the Lick survey,
which established the power-law form of the correlation function
at separations $\ls 10 \mpcoh$ ($h\equiv H_0/100 \kmsmpc$).
The following years have yielded a succession of larger
and deeper redshift surveys which probe clustering well
into the linear regime at separations approaching
$100 \mpcoh$ (e.g. CfA1: Davis \& Peebles 1983;
CfA2: de Lapparent et al. 1986; IRAS QDOT: Saunders
et al. 1991; IRAS 1.2-Jy: Fisher et al. 1993; APM-Stromlo:
Loveday et al. 1992).

These later studies have concentrated on the shape of
the large-scale fluctuation spectrum, for two main reasons:
(1) the shape of the linear spectrum allows a test of fundamental
models for the generation of fluctuations, plus the modifying effects
of dark matter; (2) redshift-space distortions mean that
redshift surveys provide a highly distorted measure of
clustering at separations $\ls 1\mpcoh$.
As long as there was uncertainty over the form of large-scale
clustering, it therefore made sense to keep the focus in this area.
However, a comparison of various large-scale  measurements by Peacock
\& Dodds (1994; PD94) suggested that the shape of  
the clustering power spectrum for wavenumbers $k\ls 0.3 \hompc$
was being measured consistently by a variety of galaxy tracers.
One aim of the present paper is therefore to concentrate on the 
small-scale data, to see how extending the range of the
clustering spectrum narrows the possible interpretations of the data.

A further development which makes it timely to
concentrate on non-linear clustering is the evolution of
the galaxy correlation function. 
The angular clustering of faint galaxies has long
indicated changes in $\xi(r)$ with epoch (e.g. Efstathiou
et al. 1991;  Neuschaefer, Windhorst \& Dressler 1991; 
Couch et al. 1993; Roche et al. 1993),
but a more direct measurement has recently been made
by the Canada-France Redshift Survey (Le F\`evre et al. 1996),
which yields spatial clustering in different redshift bins up to
$z=1$. Although a variety of different
models are capable of accounting for the local clustering
data, these alternatives make different predictions for the
change in clustering with epoch;  the second aim of the
present paper is therefore to combine the CFRS measurements
with the data on clustering at $z=0$, and to ask if any
model can successfully account for both.

The plan of the paper is as follows. Section 2 assembles
the data on the real-space clustering of optical and IRAS
galaxies, and considers their relative bias and degrees
of redshift-space distortion. 
Section 3 reviews analytical methods for obtaining
nonlinear power spectra, and proposes some empirical models
which can explain the data. Section 4 explores
the predictions for the evolution of clustering in
these models, in comparison with the CFRS data.
Section 5 adds the requirement of consistency with
fluctuations in the cosmic microwave background (CMB).
Section 6 sums up the conclusions of these studies,
which suggest that a low-density open universe with
a small degree of bias is the line of least resistance.

\sec{THE CLUSTERING OF OPTICAL AND IRAS GALAXIES}

\ssec{Clustering in real space}

It is not wise to assume in advance that any
given class of galaxy traces cosmological density in an
unbiased fashion, and we should therefore look at the properties of
a variety of tracers. Two interesting cases for which the
best data exist are optically-selected galaxies and
IRAS galaxies.
Since we are interested in clustering at small
separations, it is necessary to use real-space results.
For optical galaxies, the best power-spectrum
determination is from the angular correlations of the
APM survey (Maddox et al. 1990; 1996), as deprojected by
Baugh \& Efstathiou (1993; 1994). For IRAS galaxies,
the best available result is from the cross-correlation
between the 1-in-6 QDOT survey and its parent catalogue
(Saunders, Rowan-Robinson \& Lawrence 1992). To
reduce the data to accurate estimates of the real-space power
spectra for optical and IRAS galaxies requires a number of
steps, which are detailed below.
The resulting power spectra are shown in Figure 1, where
we follow PD94 in using a dimensionless notation for
the power spectrum: $\Delta^2$ is the
contribution to the fractional density variance
per bin of $\ln k$. In the convention of
Peebles (1980), this is
$$
\Delta^2(k)\equiv{{\rm d}\sigma^2\over {\rm d}\ln k} ={V\over (2\pi)^3}
\, 4\pi \,k^3\, |\delta_k|^2,
$$
and the relation to the correlation function is
$$
\xi(r)=\int \Delta^2\; {dk\over k}\; {\sin kr\over kr}.
$$

\ig{1}{1}
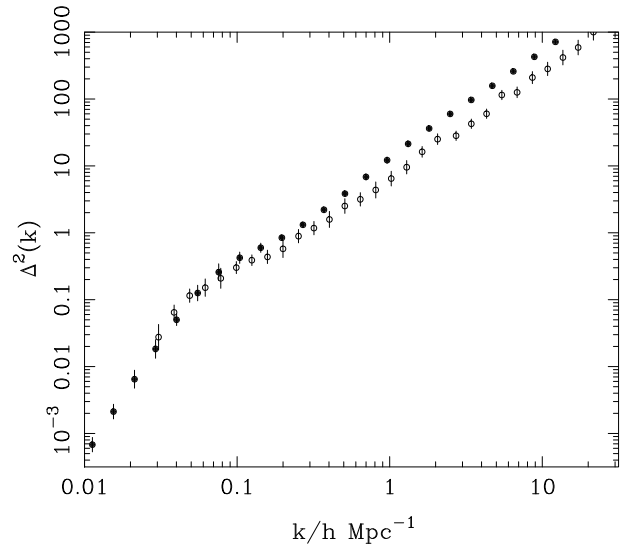
{The real-space power spectra of optically-selected
APM galaxies (solid circles) and IRAS galaxies (open circles),
obtained from the references given in the text.
Except possible on the very largest scales, IRAS
galaxies show weaker clustering, consistent with their
suppression in high-density regions relative to optical galaxies.}

A dominant topic in PD94 was how to correct redshift-survey
data for the distorting effects of peculiar velocities on the
radial coordinate; however, a variety of projection schemes
exist whereby the real-space correlations may be obtained
in an unbiased manner.
Saunders et al. (1992) used the angular cross-correlation between
the QDOT redshift survey and its (larger) parent catalogue
to obtain the projected correlation function
$$
\Xi(r)=\int_{-\infty}^\infty \xi[(r^2+x^2)^{1/2}]\; dx
= 2\int_r^\infty \xi(y)\; {y\; dy\over \sqrt{y^2-r^2}}.
$$
At first sight, it is not very attractive to
use this to infer the power spectrum, because the window
function involved is extremely broad:
$$
{1\over r} \Xi(r)= \int \Delta^2(k)\; {dk\over k}\;
\left[ {\pi\over kr}\, J_0(kr) \right].
$$
Fortunately, Saunders et al. inverted the integral relation
between $\Xi$ and $\xi$ and then converted $\xi$ to a
variance in cubical cells of side $\ell$, which is readily converted to
the power at an effective wavenumber,  as discussed by Peacock (1991):
$$
\sigma^2(\ell) = \Delta^2(k_{\rm eff}).
$$
For power-law spectra, $\Delta^2\propto k^{n+3}$, with $-2\ls n \ls 0$,
$\sigma^2(\ell)$ approximately
measures $\Delta^2(2.4/\ell)$, and the effective
wavenumber can be determined even more precisely
by performing the integral for $\sigma^2(\ell)$ over a smoothly-curving
spectrum of the observed form.

The plot in Saunders et al. (1992) which contains $\sigma^2(\ell)$ does
not extend to very small scales, but we can recover the
power spectrum in this regime because the small-scale correlations
are very close to a single power law.
Consider the projected correlations and power spectra
for a correlation function of the form $\xi=[r/r_0]^{-\gamma}$:
$$
{1\over r} \Xi(r)= {\Gamma(1/2) \Gamma([\gamma-1]/2)
\over \Gamma(\gamma/2) }\; [r/r_0]^{-\gamma}.
$$
$$
\Delta^2(k)={2\over \pi}\, \Gamma(2-\gamma)\; \sin{(2-\gamma)\pi\over 2}
\; [kr_0]^\gamma.
$$
These relations suggest an equivalent wavenumber 
$\Xi(r)/r=\Delta^2(k_{\rm eff})$, where $k_{\rm eff}=3.055/r$ for
the $\gamma=1.57$ slope characteristic of the small-scale IRAS 
correlation function. 
The IRAS power spectrum from the Saunders et al.
(1992) paper is therefore based on this method for
$k\gs 1\hompc$ and on $\sigma^2(\ell)$ for larger scales;
the two techniques match very well where they join.

The real-space power spectrum for optically-selected galaxies
is available in a much more direct form from 
the work of Baugh \& Efstathiou (1993, 1994).
They inverted the integral equations relating the angular
correlation function and the angular power spectrum to
the spatial power spectrum. The results from the first of
these papers were used in PD94. Before using them here,
however, it is advisable to make one small correction.
The Baugh \& Efstathiou formalism incorporates evolution
of clustering with redshift, but their standard results
are quoted assuming no evolution. Since the median redshift
of the $17<B<20$ samples they analyze is close to 0.2, this
can lead to a significant under-estimate of the zero-redshift
correlations. This can be checked against the recent results of
the APM-Stromlo $B<17$ redshift survey team. Loveday et al.
(1995) applied the projection technique of Saunders et al. (1992)
to deduce that $\xi=[r/5.1\, h^{-1}{\rm Mpc}]^{-1.71}$
for optically selected galaxies.
Fourier transforming the Baugh \& Efstathiou power spectrum
yields a correlation function which is uniformly lower than this  by
a factor of approximately 0.8 over to range 1 -- 10 $h^{-1}{\rm Mpc}$,
and so we have re-scaled the power spectrum to correct for this factor.
The results produced in this way are essentially identical to
those given in Maddox et al. (1996), where the deprojection was
performed allowing for evolution. The errors in Baugh \& Efstathiou (1993)
are smaller on large scales than those given by Maddox et al. (1996), even though
both are supposedly based on field-to-field scatter. To be
conservative, the larger errors were adopted.

\ssec{Optical-to-IRAS bias}

Figure 1 shows the well-known result that IRAS galaxies
cluster less strongly than optical galaxies. At least one of these classes
of galaxy must therefore be biased with respect to the mass.
The {\it relative\/} bias as a function of scale can be defined as
the square root of the ratio of the power spectra. This is
shown in Figure 2. For the smallest scales, the relative
bias is about 1.5, but this declines with scale, reaching a
plateau at $b_{\rm opt}/b_{\ss IRAS}\simeq 1.1$ around
$k\simeq 0.1 \hompc$, before declining rather abruptly
below 1 on larger scales. For
$k\ls 0.05 \hompc$, according to these data, IRAS galaxies
appear to be more clustered than optical galaxies. 
This is a suspicious feature, both for its sudden onset
and because the known suppression of IRAS galaxies in
high-density regions would lead us to expect IRAS
galaxies to be less strongly clustered on all scales.
Apart from this last feature, the striking feature
of the plot is the
slowly-varying nature of the relative bias, and it
would be useful to describe this by a fitting formula.

\ig{2}{1}
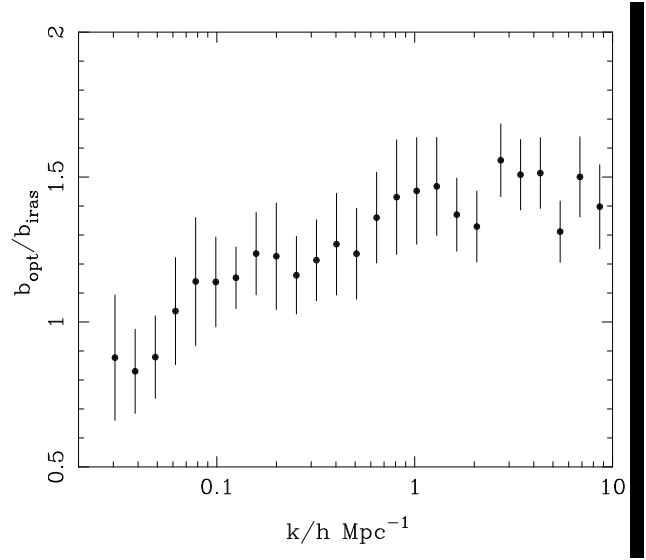
{The relative bias of optical galaxies relative to IRAS
galaxies as a function of wavenumber, obtained from the
square root of the ratio of the power spectra in Fig. 1.
The large-scale bias takes values between 1.1 and 1.2 over
a decade in scale, $0.06 \ls k \ls 0.6 \hompc$, before
breaking sharply to values $<1$ on larger scales.
}

A reasonable form to try is something which is a combination
of linear bias and a non-linear response for $\Delta^2\gs 1$:
$$
1+\Delta^2_{\rm opt}= [1+b_1\,\Delta^2_{\ss IRAS}]^{b_2}.
$$
Mann, Peacock \& Heavens (1996) show that this is a
practical form for describing many empirically reasonable forms of bias.
Local modifications of $N$-body density fields such
as suppressing particles in low-density voids or
weighting up particles in clusters produce a scale
dependence which can be fitted by this form. The behaviour is always
monotonically increasing towards small scales, as
proved on rather general grounds for all local bias
schemes by Coles (1993). The scale dependence can often
be relatively weak, as is the case here, and the observed
relative bias of optical and IRAS galaxies is therefore
very much the sort of thing that would be expected
from the known enhancement of optical galaxies relative to IRAS galaxies in 
high-density regions (Strauss et al. 1992).

The best-fitting parameters in this formula are 
$$
\eqalign{
b_1 &= 1.17\pm 0.10 \cr
b_2 &= 1.13\pm 0.01, \cr
}
$$ 
so that the linear bias parameter in the long-wavelength
regime would be
$$
b=[b_1\,b_2]^{1/2} = 1.15 \pm 0.05.
$$
Much the same result is obtained whether the whole
range of the data is used, or whether the large-scale
points with $b<1$ are omitted. Because the relative
bias is nearly a constant in the region of 1.2 over a large
range of $k$, this says that it is indeed implausible that the
large-scale points with $b<1$ can be correct.

\ig{3}{1}
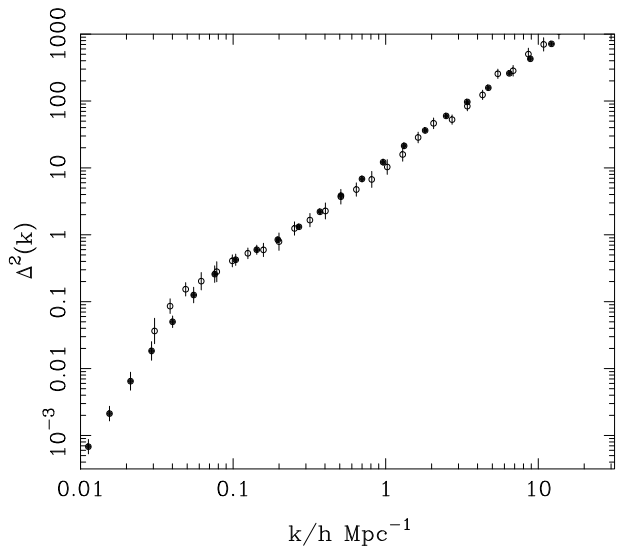
{The real-space power spectrum of optical galaxies as
estimated directly from the APM data of Fig. 1 (filled circles)
and via a smoothly-varying scaling of the IRAS data from
the same Figure. There is a good degree of unanimity concerning
the inflection around $k\simeq 0.1\hompc$, but the
two estimates diverge for $k\ls 0.06\hompc$.
}

Figure 3 shows the results of scaling the IRAS data to the
optical by this formula. The agreement in shape is
outstandingly good for all wavenumbers $k\gs 0.06 \hompc$.
Of particular interest is the inflection around
$k=0.1 \hompc$. This had been pointed out for optical
galaxies by Baugh \& Efstathiou (1993), and now appears to be
confirmed by the IRAS data -- which are derived from a
more local volume than the deeper APM data. The power spectrum is
of the form of an almost exact power law for $k>0.15 \hompc$,
but flattens quite abruptly for smaller wavenumbers.
As discussed above,  there is clear disagreement between the
optical and scaled IRAS data for $k\ls 0.05 \hompc$; rather than being
surprised by this, we should consider it remarkable that the
data agree so well to scales this large. Real-space measurements
obtained by deprojection will always be prone to systematic
errors on large scales, where the true small spatial signal
can easily become swamped in projection. To see which
determination to believe (if either) we should turn to
clustering measurements in redshift space.

\ssec{Clustering in redshift space}

There are two sets of large-scale power-spectrum
data for optical galaxies: from the CfA2 survey
(Vogeley et al. 1992) and the APM-Stromlo
survey (Loveday et al. 1992). In the latter case, the power spectrum
has been converted from
cell-count variances as in PD94, but this agrees very well
with the recent direct calculation by Tadros \& Efstathiou (1996). 
For IRAS galaxies, separate determinations have been published by
Fisher et al. (1993) and Feldman, Kaiser \& Peacock (1994).
However, it is probably better to use the result of Tadros \&
Efstathiou (1995), which combined the 1.2-Jy and QDOT datasets.
Figures 4a and 4b compare these redshift-space power
spectra with the real-space data from the previous Section,
and different behaviour is immediately apparent.
For optical galaxies, the redshift-space data lie
above the real-space results at small $k$, but fall
below them at large $k$. This is qualitatively the
behaviour that would be expected: on large scales
apparent 3D clustering is enhanced by coherent infall
(Kaiser 1987), whereas on small scales it is reduced
by the smearing of `fingers of God' due to virialized
random velocities. This trend is not really seen in
IRAS galaxies; on small scales, the real-space power
is the larger as expected, but on large scales
the IRAS real-space measurements exceeds the
redshift-space data by almost a factor of 2.

\ig{4}{2}
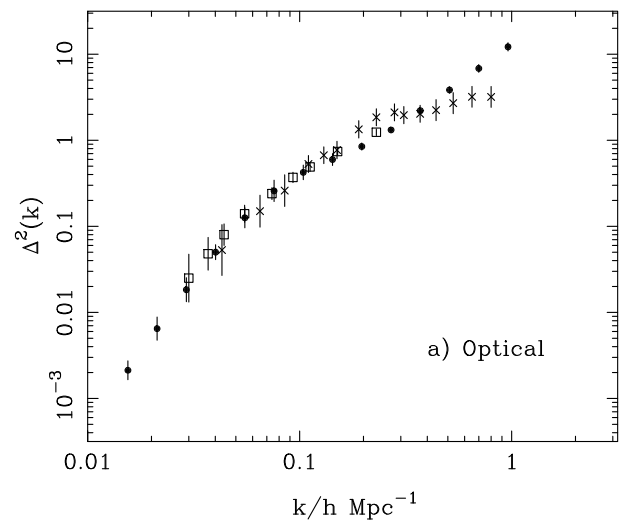
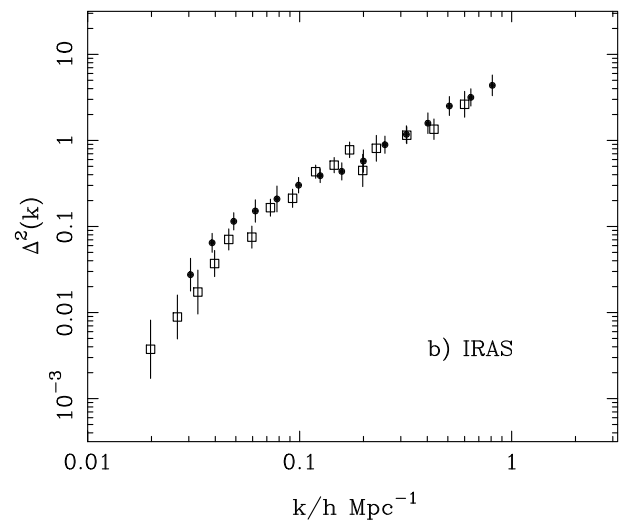
{Comparison between the real-space power spectra of Fig. 1
and redshift-space data. The real-space data are shown as
solid circles in both panels. Fig. 4a shows redshift-space
results from the APM-Stromlo survey (open squares)
and the CfA2 survey (crosses). Fig. 4b shows the redshift-space
results from the combined 1.2-Jy and QDOT surveys as open squares.
For optical galaxies, the redshift-space results are higher
on large scales, as expected from peculiar-velocity distortions.
The opposite is true for IRAS galaxies, suggesting that the
largest-scale real-space points have been overestimated.}

This is clear evidence that the
largest-scale IRAS real-space data are too high;
the three points with $k < 0.05 \hompc$ will therefore
be omitted from all subsequent analysis.
For optical galaxies, Figure 4a suggests that the APM deprojection has
worked extremely well out to the largest scale
for which it is possible to compare with the
shape of the redshift-space spectrum ($k\simeq 0.03 \hompc$).
The data considered by PD94 included other tracers on
these scales, such as Abell clusters, which showed a
large-scale slope consistent with the APM determination
out to slightly larger scales ($k\simeq 0.02 \hompc$),
and the APM data appear to be reliable within their
quoted errors up to this scale. Although results are
given to smaller wavenumbers, we will take the conservative
approach of not using them in the absence of independent
verification. The conclusion of this comparison is therefore
that the data shown in Figure 3 define the power spectrum
of optical galaxy clustering to a precision of about 20 per cent
over three decades in scale: $0.02 < k < 20 \hompc$.
With results of this verified accuracy now available, it
should be possible to reach well-constrained conclusions
about allowed models.

\ssec{Redshift-space distortions}

Before leaving this subject, it is interesting to look at
the redshift-space distortions in a little more detail, since
the size of the effect is sensitive to $\Omega$. PD94 gave
an analytical approximation for the distorting effects of
peculiar velocities on power spectra, which expresses the
power ratio as a product of the large-scale Kaiser factor
and a small-scale damping term:
$$
{P_{\rm redshift} \over P_{\rm real}} =
[1+\beta \mu^2]^2\; D[k\mu\sigma_p];
$$
the power spectrum is anisotropic ($\mu={\bf \hat k\cdot\hat r}$),
and this expression should be  averaged over $\mu$.
The parameter describing the large-scale distortion is
$$
\beta={\Omega^{0.6}\over b};
$$
the damping term depends on the assumed small-scale velocity
distribution. For an exponential pair-wise distribution with
pair-wise rms $\sigma_p$, the functional form is
$$
D[k\mu\sigma_p]= [1+(k\mu\sigma_p)^2/2]^{-1}
$$
(Ballinger, Peacock \& Heavens 1996). For describing the
effect on the power spectrum, we need $\sigma_p$ in length
units rather than velocity (i.e. $\sigma_p \rightarrow
\sigma_p/H_0$). From measurements of pairwise velocities,
sensible values for $\sigma_p$ will lie in the
range 3 -- $4\mpcoh$ (Davis \& Peebles 1983; Mo, Jing \& B\"orner
1993; Fisher et al. 1994).
The formula for the redshift-space power spectrum
ignores quasi-linear modifications,
and this assumption has been questioned by Fisher \& Nusser
(1996). However the above relation only amounts to a
linear boost of power at small $k$, with a 
reduction in power at small scales that depends quadratically
on $k$. If the expression is not
pushed to too small scales, and if we are not too interested in
the interpretation of the exact numerical size of the damping term, then this
two-parameter formula should always be able to describe the data well.

Allowing the parameters $\beta$ and $\sigma_p$
to float, a maximum-likelihood fit can be performed between
the real-space and redshift-space data (omitting the three
largest-scale IRAS points). The real-space data, having smaller
errors, are used to interpolate values at the $k$ values of
the redshift-space data, and the ratio is compared with the model.
The resulting contours are shown in Figure 5.
The optical results are far better constrained than the
IRAS, reflecting the restricted range of $k$ in the latter case,
and there is a very clear preferred model:
$$
\eqalign{
&\beta_{\rm opt} \simeq 0.40 \cr
&\sigma_p \simeq 3.5 \mpcoh. \cr
}
$$
Note that the value of $\sigma_p$ is very reasonable, 
despite the above caveats. 

From the earlier discussion of relative bias, we
would expect $\beta$ for IRAS galaxies to be a factor
1.15 larger than the optical figure. If anything, the IRAS plot appears to
favour smaller figures, but probably not too much
confidence should be placed in this, because of the
demonstrated problem with the large-scale real-space data.
The worst points have been removed, but some residual
systematics may remain in the retained points of largest
scale, and these are the ones which give the 
main signal for $\beta$.

\ig{5}{2}
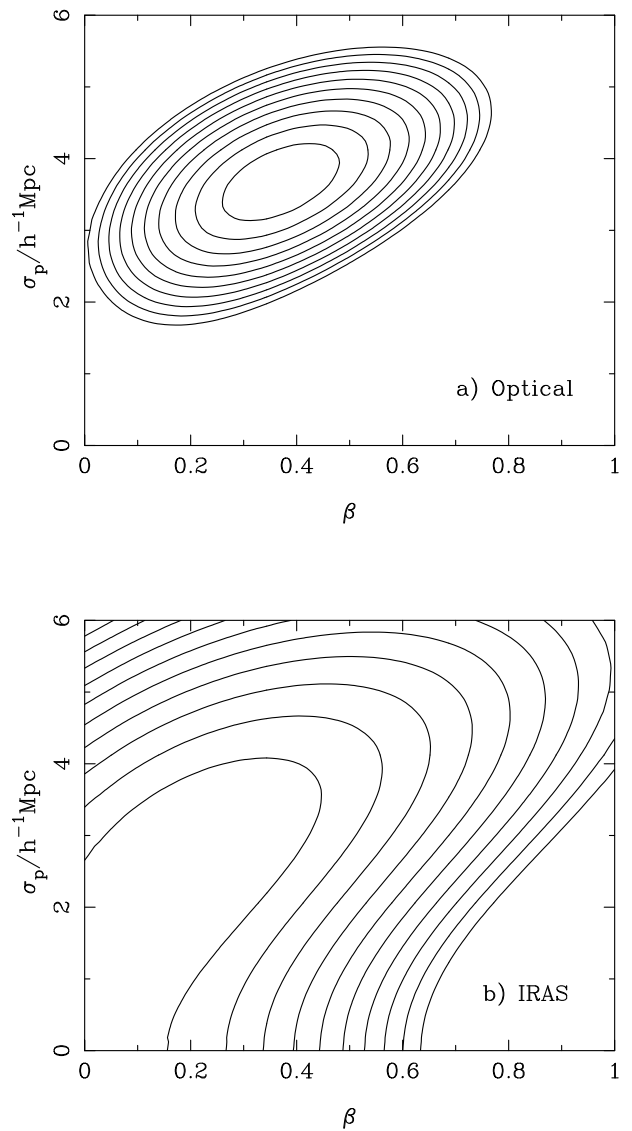
{Likelihood contours (interval 0.5 in $\ln{\cal L}$) from
fitting a model with large-scale Kaiser distortion plus
finger-of-God damping to the data of Fig. 4. The parameter
$\sigma_p$ is the pairwise velocity dispersion of galaxies
divided by $H_0$, and thus expressed in units of $\mpcoh$.
Fig. 5a shows the results for optical galaxies; Fig. 5b
shows IRAS galaxies with the three largest-scale real-space
points omitted. Note that $\beta_{\ss IRAS}
\simeq 1.15 \beta_{\rm opt}$ is expected. Even allowing for
possible systematics in the analysis, it is very hard to see how
$\beta=1$ can be allowed.}

How seriously should the low value of $\beta_{\rm opt}$ be taken? 
A good reason for taking this analysis cautiously is that
the statistical treatment of power-spectrum errors is
rather simplistic. The published data points have been treated
as independent, whereas it is inevitable that they will
be correlated to some extent (Feldman et al. 1994);
a more complete analysis would have to employ the full
covariance matrix. Nevertheless, the formal
error in $\beta_{\rm opt}$ from this analysis is very small (about 0.05), 
and it is more likely that we should be worrying about systematics.
The applicability of the Kaiser factor has
been adequately tested in simulations, so the main concern is
whether the amplitudes of either the real- or redshift-space
data could be incorrect.
The averaged linear boost from the Kaiser factor is
$$
{P_{\rm redshift}\over P_{\rm real}} =
1+2\beta/3 + \beta^2/5;
$$
for $\beta=1$ this is a factor $28/15$ as
against 1.30 for $\beta=0.4$. $\beta=1$ would
therefore be allowed if the real-space power
was overestimated by a factor 1.43, or the
redshift-space data too low by a similar amount.
It is implausible that such a large factor can be
found. The APM data were scaled by a factor of
1.25 to allow for the effects of evolution (see above).
In detail, this number was obtained from a comparison
with the amplitude of the real-space clustering in
the APM-Stromlo redshift survey. Loveday et al. (1995)
quote a scale-length of $r_0=5.1\pm0.2\mpcoh$, or an
rms uncertainty in power of 7 per cent -- much smaller than
the error needed to allow $\beta=1$.
The other worry is that different surveys may
sample galaxies of differing luminosities and thus
obtain a variety of clustering signals. Indeed,
Loveday et al. (1995) suggested that galaxies
with $ -19 < M_B < -15$  were clustered a factor roughly 2
more weakly on large scales than were more luminous galaxies.
Is it possible that such an effect has reduced the 
amplitude of the optical redshift-space data?
Certainly, the APM angular data refer to galaxies several
magnitudes fainter than the redshift surveys. The real-space
clustering signal therefore comes from a larger volume,
but it is not likely to be due to galaxies of significantly
different luminosities. Without applied redshift limits, the
selection function of magnitude-limited galaxy surveys peaks for luminosities
around $L^*$, and indeed the scaling tests for $w(\theta)$
as a function of depth applied by Maddox et al. (1990; 1996)
leave little room for a depth-dependent effect.
We therefore assume that the normalization uncertainty of 7 per cent
in power is the principal contributor to the systematic error
in $\beta_{\rm opt}$. Added in quadrature to the statistical uncertainty,
this gives the minimal error bar
$$
\beta_{\rm opt}=0.40 \pm 0.12.
$$
This uncertainty is approximately four times smaller than that
obtained from a fuller analysis of the APM-Stromlo
survey by Tadros \& Efstathiou (1996),
but it is not necessarily unrealistic. Tadros \& Efstathiou
considered $k\ls0.1\hompc$ only, which increases the noise,
and we have considered here both the APM-Stromlo and CfA2 surveys.

In any case, a figure of $\beta_{\rm opt}\simeq 0.4$ is certainly consistent
with a number of recent determinations from measurements of
clustering anisotropy within a survey. 
Ratcliffe et al. (1996) find $\beta_{\rm opt}=0.55\pm0.12$
from the Durham/UKST survey, and Loveday et al. (1996) obtain
$\beta_{\rm opt}=0.48\pm0.12$ from the APM/Stromlo survey.
From the IRAS 1.2-Jy survey, Cole, Fisher \& Weinberg (1995)
obtain $\beta_{\ss IRAS}=0.52\pm0.13$.
This convergence on a value of around $\beta=0.5$ is in very serious
disagreement with the values around unity inferred from
the comparison of peculiar velocities with the galaxy
density field (Dekel 1994), and it is of the greatest
importance that the discrepancy be resolved. The impression
from the present work is that $\beta=1$ would require
gross errors in clustering measurements, and it seems
implausible that this can be the correct value.
The lower values suggest that $\Omega\simeq 0.2$ -- 0.3
if optical galaxies trace mass, or require $b\simeq 2$
in an Einstein-de Sitter universe.

\sec{NONLINEAR MASS POWER SPECTRA}

\ssec{Analytical method}

In order to compare the clustering data with theoretical
models, it is necessary to have an accurate means of predicting
the effects of nonlinear growth on a given linear power spectrum.
This was made possible by the insight of Hamilton et al. (1991),
who suggested a scaling formula for the growth of correlation
functions. PD94 extended this method to work with power
spectra and also to universes with $\Omega\ne 1$. The original
suggestion of Hamilton et al. was that their scaling formula was
spectrum independent, but Jain, Mo \& White (1995) showed that
this was not true. Peacock \& Dodds (1996; PD96) give an
improved version of the PD94 method which takes this into account,
allowing the nonlinear spectrum that results from any 
smoothly-varying linear spectrum to be calculated.
The method may be summarized as follows.

The nonlinear spectrum is a function of the linear spectrum at
a smaller linear wavenumber:
$$
\Delta^2_{\ss NL}(k_{\ss NL}) = f_{\ss NL}[\Delta^2_{\ss L}(k_{\ss L})],
$$
$$
k_{\ss L} = [1+\Delta^2_{\ss NL}(k_{\ss NL})]^{-1/3} k_{\ss NL}.
$$
PD96 give the following fitting formula for
the nonlinear function:
$$
f_{\ss NL}(x) =x \; \left[{ 1+B\beta x +[A x]^{\alpha\beta} \over
1 + ([A x]^\alpha g^3(\Omega)/[V x^{1/2}])^\beta}\right]^{1/\beta}.
$$
In this expression, $B$ describes a second-order deviation from linear
growth; $A$ and $\alpha$ parameterise the power-law which
dominates the function in the quasilinear regime; $V$ is
the virialization parameter which gives the amplitude of the
$f_{\ss NL}(x) \propto x^{3/2}$ asymptote
(where the behaviour enters the `stable clustering' limit); $\beta$ softens
the transition between these regimes.
The parameters and their dependence on spectrum are
$$
A=0.482\,(1+n/3)^{-0.947}
$$
$$
B=0.226\,(1+n/3)^{-1.778}
$$
$$
\alpha=3.310\,(1+n/3)^{-0.244}
$$
$$
\beta=0.862\,(1+n/3)^{-0.287}
$$
$$
V=11.55\,(1+n/3)^{-0.423}.
$$
For linear spectra which are not a power-law, particularly
the CDM model, PD96 suggested that
a tangent spectral index as a function of linear wavenumber
should be used:
$$
n_{\rm eff}(k_{\ss L})\equiv {d\ln P \over d \ln k}(k=k_{\ss L}/2).
$$ 
The factor of 2 shift to smaller $k$ is required 
because the tangent power-law at $k_{\ss L}$ 
overestimates (underestimates) the total degree of nonlinearity
for curved spectra where $n_{\rm eff}$ decreases
(increases) as $k$ increases;
the results are not greatly sensitive to the exact shift.
This prescription is able to predict the nonlinear evolution
of power-law and CDM spectra up to $\Delta^2\simeq 10^3$
with an rms precision of about 7 per cent.

Note that the cosmological model is present in the fitting formula
only through the growth factor $g$, which governs the
amplitude of the virialized portion of the spectrum.
According to Carroll, Press \& Turner (1992), the
growth factor
may be approximated almost exactly by
$$
g(\Omega) ={{5}\over{2}}\Omega_{\rm m}\left[\Omega_{\rm m}^{4/7}-\Omega_{\rm v}+
(1+\Omega_{\rm m}/2)(1+\Omega_{\rm v}/70)\right]^{-1},
$$
where ($m$) and vacuum ($v$) contributions to the density 
parameter are distinguished; $\Omega$ without a subscript
generally means $\Omega_{\rm m}$.
When considering non-zero vacuum energy, it is usual to restrict
attention to spatially flat models only. Models with
$\Omega<1$ thus come in two varieties: open ($\Omega_v=0$)
and flat ($\Omega_v=1-\Omega_m$).

\ssec{CDM models and normalization of power spectra}

As an illustration of these techniques in action, it is
interesting to consider the CDM spectrum, which is a
popular choice for the empirical description of the
power spectrum, being a smoothly-curving form parameterised
by a normalization $\sigma_8$ and a shape $\Gamma^*$.
The CDM power
spectrum is $\Delta^2(k)\propto k^{n+3}T_k^2$; usually
$n=1$ is adopted for the primordial spectrum.
We shall use the BBKS approximation for the transfer
function (which unfortunately suffered a misprint in PD94):
$$
\eqalign{
T_k= & {\ln(1+2.34q)\over 2.34q}  \times \cr
     & [1+3.89q+(16.1q)^2 +(5.46q)^3 +(6.71q)^4]^{-1/4}, \cr
}
$$
where $q\equiv (k/h\; {\rm Mpc}^{-1})/ \Gamma^*$.
The shape parameter $\Gamma^*$ would be equal to $\Omega h$
in a model with zero baryon content. PD94 showed that there
was a significant shift as a function of $\Omega_{\ss B}$,
which was generalized to models with $\Omega\ne 1$ by
Sugiyama (1995):
$$
\Gamma^*=\Omega h\, \exp[-\Omega_{\ss B}(1+1/\Omega)].
$$
Because of this scaling, $\Gamma^*=0.94 \Gamma$, where
$\Gamma$ is the shape parameter defined by
Efstathiou, Bond \& White (1992).

\ig{6}{1}
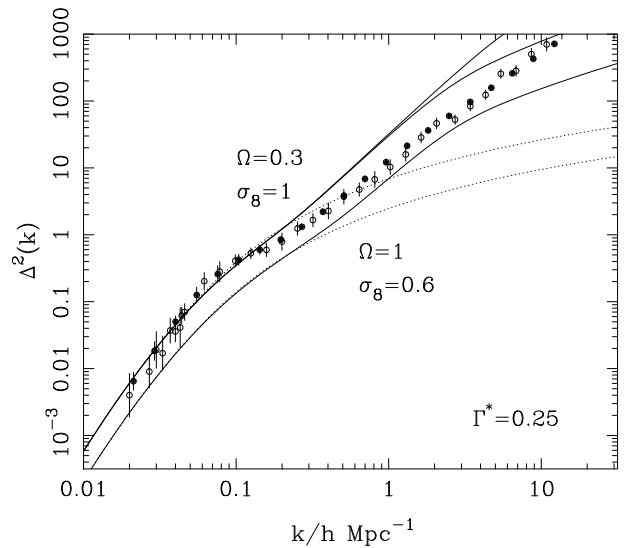
{The optical-galaxy power spectrum of Fig. 3 with the
discrepant scaled IRAS points 
omitted (APM data are shown as solid points, scaled IRAS
data as open circles for $k > 0.05 \hompc$).
Also shown as open circles for smaller wavenumbers are the
redshift-space data from Fig. 4, corrected assuming
$\beta_{\rm opt}=0.4$, $\sigma_p=3.5 \mpcoh$.
CDM models with
$\Gamma^*=0.25$ and $\sigma_8=0.6$ \& 1 are shown. Linear
spectra are shown dotted, with nonlinear spectra shown solid.
For the lower normalization, $\Omega=1$ is assumed; for
the higher normalization, $\Omega=0.3$ open and flat models
are shown (open being the higher). The unbiased models
greatly exceed the small-scale data; the $\Omega=1$ model
would require bias which was a strongly non-monotonic function of scale.}

The normalization is specified by $\sigma_8$: the linear-theory rms
density contrast when averaged over spheres of radius $8\mpcoh$:
$$
\sigma_R^2 =\int \Delta^2(k)\;{{\rm d}k\over k}
\;{9\over (kR)^6}[\sin kR -kR\cos kR]^2.
$$
$\sigma_R^2$ is just $\Delta^2(k)$ at some effective wavenumber,
which is very well fitted by
$$
k_{\rm eff}/ \hompc = 
0.172 + 0.011 [\ln (\Gamma^*/0.34) ]^2.
$$

What normalization is appropriate? A strong constraint
comes from the abundance of massive clusters of galaxies,
which is exponentially sensitive to the normalization,
in a way that depends very little on spectrum.
White, Efstathiou \& Frenk (1993) argue that this gives
$$
\sigma_8 =0.57\;\Omega^{-0.55},
$$
to a tolerance of roughly 10 per cent. They consider
spatially flat models only, but the scaling for open
models is likely to be very similar.
Notice that the content of this equation is very
similar to the effect of redshift-space distortions.
The apparent (but non-linear) value of $b_{\rm opt}\sigma_8$
is about 1, so the cluster constraint implies
$\beta_{\rm opt}\simeq 0.57$. This is a further reason for
thinking that there must be a serious problem with the
values of $\beta\simeq 1$ deduced from velocity-field studies.
This also says that the cluster abundance constraint
gives a very similar value for the normalization to that
obtained from small-scale pairwise velocity dispersions,
since they scale with $\Omega$ in almost the identical way.
Taken literally, the White et al. normalization would
imply that optically-selected galaxies are antibiased for
$\Omega \ls 0.3$, whereas $M/L$ ratios in clusters would
prefer $\Omega \simeq 0.15$ -- 0.2 in the absence of bias.
We shall ignore this slight uncertainty and treat low-density
models under the assumption that optically-selected galaxies
trace mass as a first approximation: $\sigma_8=1$. For Einstein-de Sitter
models, however, this assumption would make no sense and
a lower normalization must be adopted; we take $\sigma_8=0.6$.
Small variations on these assumptions do not change the conclusions.

\ig{7}{2}
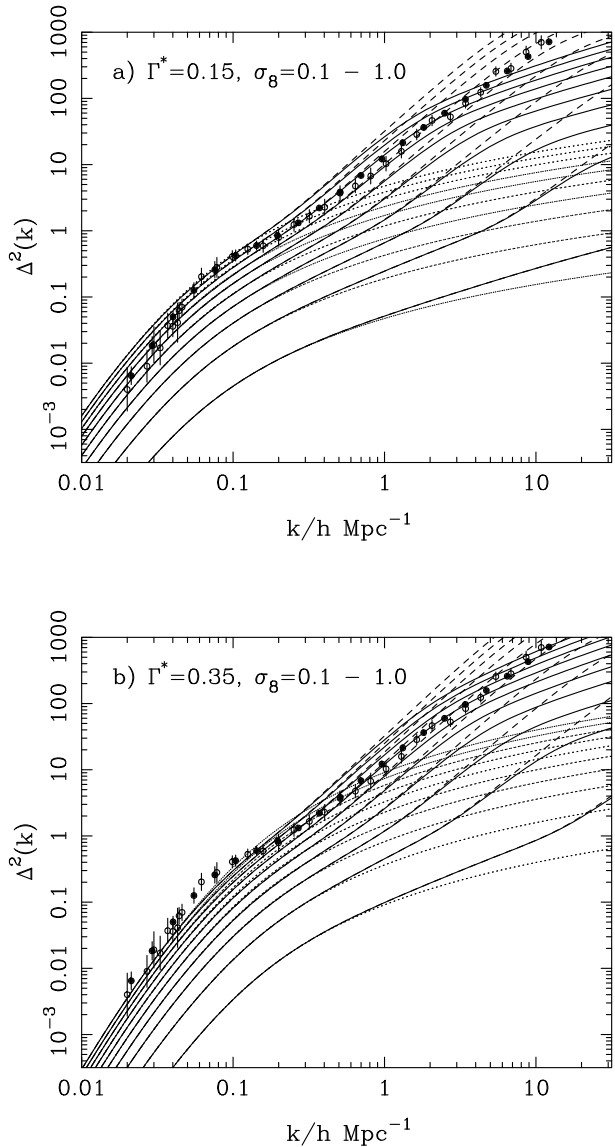
{An extended form of Fig. 6, showing CDM spectra with
a) $\Gamma^*=0.15$, b) $\Gamma^*=0.35$, for $\sigma_8$
values between 0.1 and 1, in steps of 0.1.
Linear spectra are dotted; the solid lines show 
nonlinear spectra for $\Omega=1$, 
and the dashed lines show an open $\Omega=0.3$ model.}

Figure 6 shows the predictions of CDM models with these normalizations
and the shape parameter $\Gamma^*=0.25$ claimed by PD94
to fit the shape of large-scale structure very well.
The Einstein-de Sitter model and $\Omega=0.3$ models 
(open and flat) are shown, and none fare very well.
As expected, the low-density models follow the low-$k$
clustering data rather well, but both open and flat models
start to predict too much small-scale power for $k\gs 0.2 \hompc$,
and they do not produce much of an inflection around
$k\simeq 0.1 \hompc$. The open and flat models fare equally
badly in this respect, only starting to part company at
$k\gs 1 \hompc$. Changing $\Omega$ would only change the
predicted spectrum in this regime, and so would not cure
the problem. This tendency for low-density unbiased CDM
models to overpredict the small-scale power was noted
in the $\Gamma^*=0.2$ case by Efstathiou,  Sutherland \&
Maddox (1990), and recently rediscovered by Klypin et al. (1996).
Because of its lower normalization, the $\Omega=1$ model
at least does not exceed the data, but it still presents
serious problems. The implied bias is an oscillating
function of scale: it has a local maximum around
$k\simeq 0.1\hompc$ (again, no inflection is predicted), and a
local minimum around $k\simeq 3\hompc$. No local bias
scheme could give this non-monotonic behaviour (Coles 1993). 

These problems are general, and not specific to the
exact value of $\Gamma^*$ used, or the tilt adopted.
For models with $n\ne 1$, the results over the range
of $k$ considered are indistinguishable from those with
$n=1$ and a different value of $\Gamma^*$.
Figure 7 shows a grid of models with varying $\sigma_8$
for $\Gamma^*=0.15$ and 0.35, and they show that
the same kind of problems persist:

\item{(1)} Unbiased models with $\sigma_8\simeq 1$ over-produce
small-scale power. The correct level of power around
$k=1 \hompc$ requires $\sigma_8\simeq 0.7$, but the large-scale
data are then very badly fitted, whatever value of $\Gamma^*$
is assumed.

\item{(2)} No model reproduces the inflection seen 
around $k\simeq 0.1 \hompc$.

\item{(3)} Only open low-density models produce a small-scale
power spectrum which resembles the observed power law.
Both $\Omega=1$ models and flat low-density models
reach the stable clustering regime and turn to a
flatter slope rather abruptly at $\Delta^2\simeq 50$ -- 100.
These models need a bias scheme capable of
generating a sharp kink at just the required point.

The reason PD94 were not able to reach these
conclusions is because of the improved 
nonlinear treatment used here, which allows
high accuracy in the predictions at $0.5 \ls k \ls 10 \hompc$.
PD94 considered only the deduced {\it linear\/}
spectrum at $\ls 0.5 \hompc$, which reduces the amplitude
of the conflict between CDM models and observation.
Comparison of nonlinear spectra is clearly the more
sensitive test, given accurate nonlinear predictions;
the nonlinear treatment of PD96 in fact gives a
slightly larger response for low-$\Gamma^*$ CDM models,
which exacerbates the difficulties for CDM.
The conclusion is therefore that the CDM spectrum
is of the wrong shape, and cannot be made to fit
the data, however its parameters are adjusted.

\ssec{Empirical linear power spectra}

What appears to be required is a linear spectrum which bends more abruptly around 
$k\simeq 0.1 \hompc$, and the inflection at this point is therefore to be
interpreted as a residual feature from the linear
spectrum. The precise linear form can be recovered by running
the nonlinear apparatus backwards, although this is
now a more complicated procedure than in PD94, because the
nonlinear response depends on the slope of the linear
spectrum. We have to proceed iteratively, assuming an
initial linear spectral slope, and then refining the
dependence of this slope on scale.

For low-density models, it is reasonable to
perform this linearization under the assumption that
the clustering of optically-selected galaxies is
not significantly biased. Models with $\Omega=1$
present greater difficulties: there is no point in
recovering a linear spectrum which would evolve
into the observed clustering when we know that the
resulting amplitude of mass fluctuations would be far too high.
Before performing the linearization, we need to allow
for bias, but this requires a specific hypothesis 
for the variation of bias with scale. Since we do not know
what this variation is, the best that can be done is to
illustrate how the linearized spectrum depends on different
assumptions for the bias. The nonlinear response is
by definition highly sensitive to the linear spectrum, so there
is at least some hope that the recovered linear spectrum will
not depend critically on assumptions about bias. We
shall consider two examples. The simplest is a constant
linear bias:
$$
\Delta^2_{\rm opt}=b^2 \Delta^2_{\rm mass};
$$
in the light of the earlier discussion, $b=1.6$ is about
the correct number. This is certainly an extreme example,
since we have seen that bias generally increases at small
scales. For an alternative model, we can take a hint from
nature and guess that the scale dependence of the bias
is similar to the form that relates optical galaxies and
IRAS galaxies:
$$
1+\Delta^2_{\rm opt}= [1+b_1 \Delta^2_{\rm mass}]^{b_2}
$$
Non-linear bias models of this type are considered in detail by 
Mann, Peacock \& Heavens (1996). 
Their work shows that a broad range of biasing prescriptions
yield a galaxy power spectrum with a relationship to the mass power
spectrum which is well fitted by such a functional form.
For IRAS/optical bias, we had $b_1\simeq b_2$, and so this
suggests $(b_1,b_2)=(1.6,1.6)$, for an overall linear bias
of 1.6. There is no guarantee that this is the exact bias
that applies to the real universe, but it will suffice 
to illustrate the effects of a realistic amount of scale-dependent bias.
Figure 8 shows the recovered linear spectra for these various
cases.

\ig{8}{2}
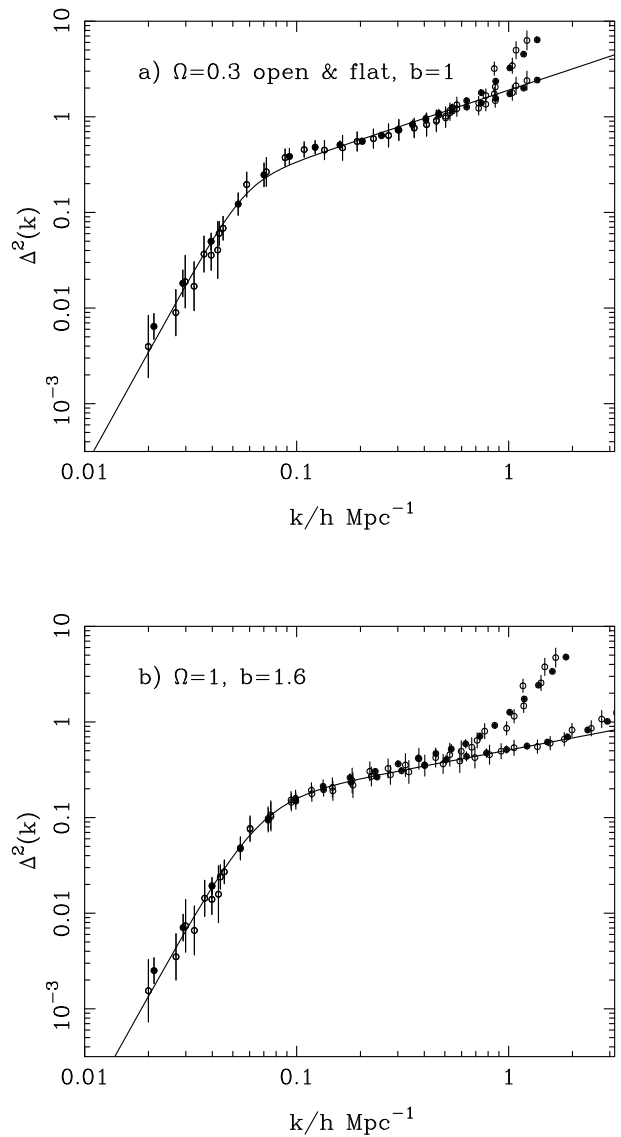
{Linearized versions of the optical power-spectrum data
from Fig. 3, with APM results shown as solid points and
scaled IRAS results as open circles.
The solid lines are the power-law fits described in the text.
Fig. 8a assumes no bias and shows results for open and flat $\Omega=0.3$ models.
These are similar on large scales, but the flat results require
an abrupt steepening on small scales.
Fig. 8b assumes $\Omega=1$ and a linear bias of $b=1.6$. If this is
independent of scale, a small-scale feature is again required.
For the more plausible scale-dependent bias described in the text,
the linear spectrum is more nearly a single power law on small scales.}

In all cases, what we see is a spectrum which breaks
rather abruptly, as expected from the discussion of the
failures of the CDM spectrum.
As hoped, the $\Omega=1$ reconstructions with different
bias assumptions differ only slightly.
For $k\ls 1\hompc$, a satisfactory empirical description of the data 
is provided by a simple transition between two power laws:
$$
\Delta^2(k)={(k/k_0)^\alpha\over [1+(k/k_1)^{(\alpha-\beta)/\delta}]^\delta},
$$
as shown in Figure 9.
For low-density unbiased models, the parameters are
$$
\eqalign{
k_0 &  = 0.42\; h\, {\rm Mpc}^{-1} \cr
k_1 &  = 0.057\; h\, {\rm Mpc}^{-1} \cr
\alpha &  = 0.74 \cr
\beta &  = 4.0 \cr
\delta & = 0.6. \cr
}
$$
A value of $\beta=4$ corresponds to a scale-invariant
spectrum at large wavelengths, whereas the effective
small-scale index is $n=-2.3$. 
Note that here and below we focus on $\Omega=0.3$ as
a representative low-density model, but almost indistinguishable
linearized results would be obtained if this was varied by up to a factor 2
(unless bias was varied as a function of $\Omega$).
For $\Omega=1$, the parameters change to
$$
\eqalign{
k_0 &  = 5.0\; h\, {\rm Mpc}^{-1} \cr
k_1 &  = 0.066\; h\, {\rm Mpc}^{-1} \cr
\alpha &  = 0.42 \cr
\beta &  = 4.0 \cr
\delta & = 1.0, \cr
}
$$
independent of the assumed
scale dependence of bias for linear wavenumbers $k\ls 0.4\hompc$.
These fits are similar in form to a model advocated by
Branchini, Guzzo \& Valdarnini (1994). On the basis of
a redshift survey of the Perseus-Pisces region, they
suggested a phenomenological power spectrum in which
the small-scale index was $n=-2.2$, with a break at
$k\simeq 0.08 \hompc$. They did not consider
the nonlinear evolution of any alternative model, but this
was an early indication that a very flat small-scale
spectrum might fit the clustering data.

It should be noted that this linearization process 
is close to being unstable,
because the nonlinear response rises very rapidly as the
linear spectrum approaches $n=-3$. The small-scale index
cannot greatly exceed $n=-2.3$, but it could be flatter.
This is an issue for the low-density models in particular,
as can be seen in Fig. 8a. The analytic fit is not
perfect, with the data lying slightly higher around
$k=0.1\hompc$ and lower around $k=0.3\hompc$.
It is not possible to achieve a better fit, since this would 
require a local $n\simeq -3$ between $k\simeq 0.1\hompc$
and $0.2\hompc$. Such a flat spectrum would suppress the required
power at $k=0.3\hompc$, resulting in a peak in the linear
spectrum at $k\simeq 0.1\hompc$. This is a fascinating
possibility, but needs further investigation;
the PD96 formulae have not been tested in the regime where
the concept of a hierarchical density field breaks down.
In any case, the errors on the data are such that the
apparent misfit in Fig. 8a is of questionable significance.
For the present, the conservative approach is to stick
with a monotonically increasing $\Delta^2(k)$, and the
linearization in Fig. 8a therefore assumes the local slope
$n(k)$ given by the analytic fit.

\ig{9}{1}
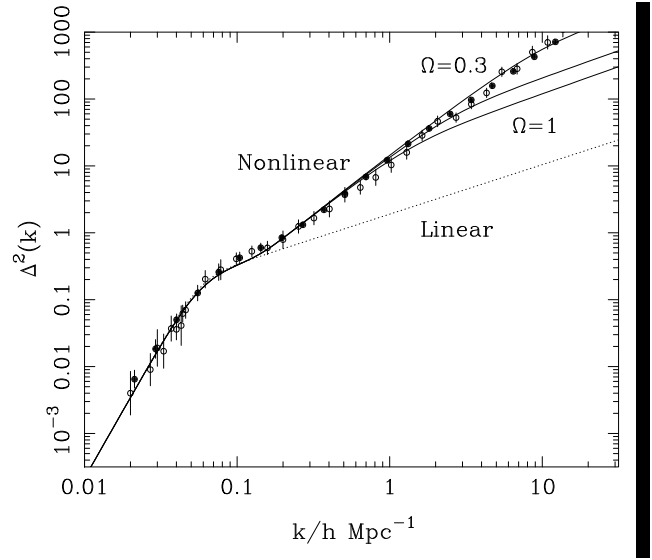
{Illustrating the fit of the two power-law
model for the linear spectrum in the case of an $\Omega=0.3$
open model. For the same linear spectrum, $\Omega=0.3$ flat and
$\Omega=1$ models produce successively more marked
flattening of the small-scale nonlinear spectrum. In these latter cases,
a small-scale feature in the linear spectrum is required
in order to keep the nonlinear spectrum steep.}

The fits work well enough at large scales, but in most
cases they fail to fit the reconstruction at the smallest
scales, and there is generally a `bump' where the power
increases quite abruptly. This can be understood with
reference to the nonlinear spectra shown in the discussion
of the evolution of the CDM spectrum. High-density and flat
models with $\Lambda\ne 0$ tend to hit the virialized
regime quite early, flattening the predicted power spectra
for $\Delta^2\gs 100$. This is not seen in the
data, and so a feature must be introduced into the
linearized spectra in order to keep the nonlinear spectrum
steep.
For open models, this feature is not required for $\Omega<0.3$.
For $\Omega=0.3$, the small-scale flattening is just starting to become
significant, and smaller values would arguably be preferred. However,
as discussed below, the higher value gives a better match to
the data on clustering evolution.
The fitting formula is easily amended to deal with
this small-scale behaviour:
$$
\Delta^2(k)={(k/k_0)^\alpha + (k/k_2)^\gamma\over 
[1+(k/k_1)^{(\alpha-\beta)/\delta}]^\delta}.
$$
A value of $\gamma=3$ ($n=0$ spectrum) is appropriate, and the
required normalization is $k_2\simeq 0.72 \hompc$ for the flat
$\Omega=0.3$ case (increasing to 0.79 for $\Omega=0.2$), 
or $k_2\simeq 0.91 \hompc$ for $\Omega=1$.
The spatial scales of interest for the 
small-scale feature, where correlations in the
range 100 -- 1000 are found, are separations of about
0.5 -- $0.2 \mpcoh$. 

The alternative explanation is that the
mass spectrum does flatten on these scales, but bias
steepens it again. This is hard to rule out, but it is
does seem a little contrived that the effects of bias
and gravitational nonlinearity should conspire to cancel
each other out in this way. The only case where such a
conspiracy is not required is when the universe is open
with $\Omega\ls 0.3$; the virialized regime then occurs at
sufficiently high overdensity that the predicted
nonlinear spectrum remains almost a single power law
up to $\Delta^2\gs 1000$, without having to introduce a
feature in the linear spectrum. This point is illustrated
in Figure 9.

\ssec{Physical models}

What is the interpretation of the linear spectrum?
We have argued that CDM models are inadequate for
describing anything but the very large-scale
portion of the spectrum, and so the temptation
to put a physical meaning to the best-fitting
shape parameter through $\Gamma^*\simeq \Omega h$
may not always be correct. Nevertheless, 
many of the alternative models produce less small-scale
power than a CDM model of a given density, so a
lower limit to $\Omega h$ may be obtained by this argument.
In this case, it is worth noting that a substantial
density is implied: $\Omega \le 0.2$ would be difficult to
reconcile with any reasonable Hubble constant.

Within the compass of models where the dark matter
is still cold and collisionless, there is only one
alternative in the literature that produces a
spectrum with the required sharp break:
isocurvature initial conditions (Efstathiou \& Bond 1986).
However, in this case the large-scale rms CMB anisotropies
would be a factor of 6 higher for a given level of
large-scale matter fluctuations, and so this possibility
must be rejected.

\begintable{1}
\leftline{{\bf Table 1.} Best-fitting parameters for physical power spectra.}
\bigskip
\centerline{\vbox{\tabskip 0.5cm
\halign{\hfil#\hfil & \hfil#\hfil & \hfil#\hfil & \hfil#\hfil \cr
Model & $\Omega$ & $\Gamma^*$ & $f_\nu$ \cr
\cr
CDM &  1  & 0.109 & 0 \cr
CDM & 0.3 & 0.090 & 0 \cr
HDM &  1  & 0.440 & 1 \cr
HDM & 0.3 & 0.429 & 1 \cr
MDM &  1  & 0.373 & 0.307 \cr
MDM & 0.3 & 0.393 & 0.335 \cr
}
}}
\endtable

Assuming that `designer inflation' and features
in the primordial spectrum are rejected, we are left 
with the alternative of modifying the matter content.
Interesting possibilities with high density are
either mixed dark matter (e.g. Holtzman 1989; 
Taylor \& Rowan-Robinson 1992; Klypin {\it et al.} 1993),
or non-Gaussian pictures such as cosmic strings + HDM,
where the lack of a detailed prediction for
the power spectrum helps ensure that the model is
not yet excluded (Albrecht \& Stebbins 1992).
It has been argued that an MDM spectrum with $\Omega h\simeq 0.5$ and
approximately 30 per cent of the density in a hot
component gives a spectral shape quite close to the empirical one.
This would be true whether or not the total density was $\Omega=1$,
so one could also consider low-density MDM models.
Pogosyan \& Starobinsky (1995) give analytic formulae for the
MDM transfer function, which they express as 
$T_{\ss MDM}=T_{\ss CDM}(k,h)D(k,h,\Omega_\nu)$, assuming $\Omega=1$.
If we remove this assumption, then the formulae can
still be used, replacing $k/h^2$ by $[k/h]/\Gamma^*$ and
$\Omega_\nu$ by $f_\nu=\Omega_\nu/\Omega$.
Note that there appears to be a misprint in the definition
of the parameter $\beta$ after their equation (3):
it should be $\beta=[5-(25-24\Omega_\nu)^{1/2}]/4$.

\ig{10}{2}
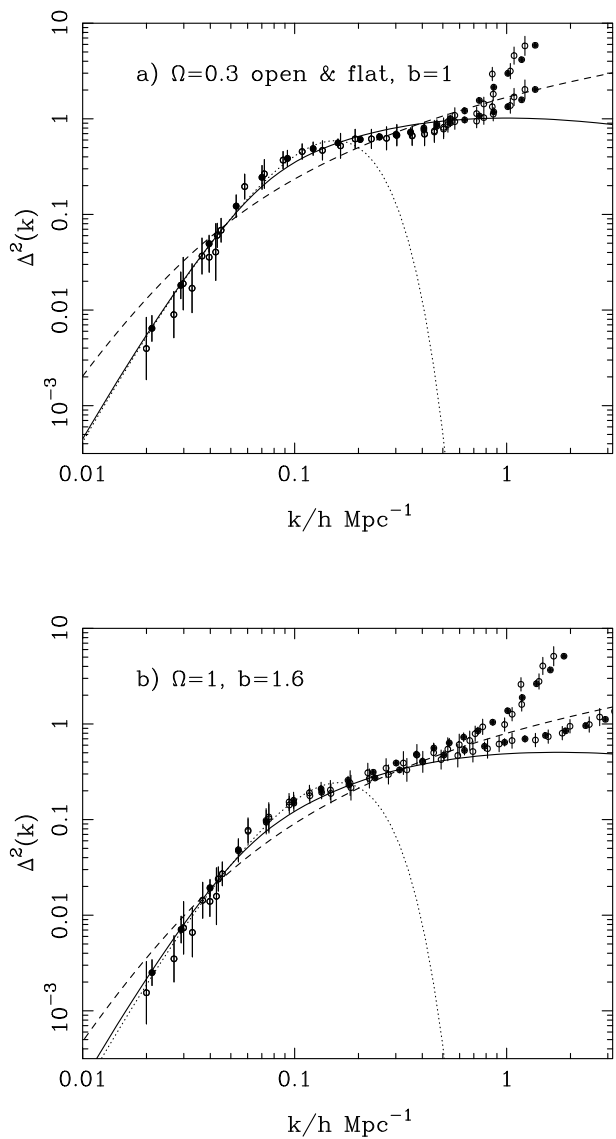
{The linearized spectrum data from Fig. 8, fitted with
CDM (dashed), HDM (dotted) and MDM (solid) models.
The best-fitting parameters are given in Table 1.
MDM spectra provide the best fit, although HDM works
well over the range to which the fit was restricted
($k<0.2\hompc$ for HDM; $k<1 \hompc$ in the other cases).}

Fig. 10 shows the fit of this 2-parameter form to the linear power data,
and contrasts this with the poor fit of CDM alone.
This plot also compares pure HDM models, restricting the fit to the
large-scale portion of the spectrum. Both HDM and MDM achieve a
satisfactory fit for $k\ls 0.2\hompc$, and both spectra
in fact make similar nonlinear predictions, underlining the
earlier discussion on the instability of the 
linearization at large $k$.
The parameters for these models are given in Table 1. 
The preferred values of $\Gamma^*$ for HDM and MDM are something of
a puzzle, since they are in the region of 0.4.
This is too large to be consistent with low-density models
for any reasonable Hubble constant, and would require
$h\simeq 0.4$ for $\Omega=1$. Many would also think that
this is an unreasonable number, but it does at least yield
a reasonable age for $\Omega=1$ models.
The Einstein-de Sitter universe 
is thus the only case where a consistent picture for the
present universe can be made
using an {\it a priori\/} physical model for the power spectrum.

However, both HDM and MDM models face problems that are
generic to any model with a very flat high-$k$ spectrum in
a high-$\Omega$ universe: difficulty in forming high-redshift
objects. These are classic problems for HDM models, but
also appear to be fatal for MDM (Mo \& Miralda-Escud\'e 1995; 
Ma \& Bertschinger 1994; Mo \& Fukugita 1996).
As shown above, high-density models also have 
difficulty in achieving a steep correlation function on small scales;
both this point and high-$z$ galaxy abundances argue for an additional
small-scale component in the linear spectrum, if $\Omega=1$.
Both these difficulties disappear in an open universe, but
it is hard to feel much enthusiasm  for the ugly combination
of MDM and $\Omega<1$. Apart from aesthetics, we have seen that
the value of $\Gamma^*$ in this model is much higher than would
be required in a universe with reasonable $h$ and $\Omega\simeq 0.3$.
In summary, all known physically-motivated models for the
clustering power spectrum face very serious difficulties. It
appears that some completely new alternative is needed.

\sec{EVOLUTION OF CLUSTERING}

Having identified a variety of routes by which the
present-day clustering of galaxies could have arisen,
the obvious way to break the degeneracy is to look at
the change of clustering with epoch, for which the
various models make quite different predictions.
For a number of years, the evolution of clustering has
been probed by means of the angular correlation function,
$w(\theta)$, and its change with magnitude limit.
Limber's equation (e.g. Peebles 1980) allows a relation
to be made between $w(\theta)$ and $\xi(r)$ once the
redshift distribution is known for the galaxies under
study. A number of studies have been carried out, of
which some of the more recent are Efstathiou et al (1991);
Neuschaefer, Windhorst \& Dressler 1991;
Couch, Boyle \& Jurcevic (1993); Roche et al. (1993).
Although there has been some debate over the interpretation
of these studies owing to the uncertain redshift distribution
at very faint limits, there has been a measure of agreement
over two conclusions: 

\item{(1)} The slope of the correlation function appears
not to alter with redshift. If $\xi(r)\propto r^{-\gamma}$,
then $\gamma$ is in the region of 1.7 at all redshifts
up to $z\simeq 1$.

\item{(2)} The rate of evolution of clustering is relatively
rapid, close to the linear-theory rate
of evolution in an Einstein-de Sitter universe: $\xi(r,z)\propto 
(1+z)^{-2}$.

\ssec{The predicted rate of evolution}

Several workers have commented that these conclusions are paradoxical.
A common model for predicting the evolution of clustering is
Peebles' stable-clustering argument. In the limit that
clusters are virialized entities of fixed proper size, 
clustering will change with epoch only because the background
density evolves as $(1+z)^3$. In proper length units, this
predicts $\xi(r_p,z)\propto (1+z)^{-3}$. In comoving units,
used everywhere else, one is then led to write
$$
\xi(r,z) = [r/r_0]^{-\gamma}\, (1+z)^{-(3-\gamma+\epsilon)},
$$
where $\epsilon=0$ is stable clustering; $\epsilon=\gamma-3$ is
constant comoving clustering; $\epsilon=\gamma-1$ is 
$\Omega=1$ linear-theory evolution. It is usually argued that
biased CDM-like universes would therefore predict $\epsilon$
significantly less than zero, whereas the data favour
$\epsilon\simeq 1$. An alternative way of looking at this is
to ask how the typical proper size of a cluster changes with
redshift: $R_p\propto (1+z)^{-\beta}$. Clearly, $\epsilon=(\gamma-3)\beta$
and so $\epsilon\sim 1$ implies that $\beta\sim -1$: clusters
must contract about as fast as the universe expands.

In short, the paradox is that galaxy clustering is required
to evolve in a way that resembles linear theory in an
$\Omega=1$ universe, keeping the same shape and changing its
amplitude rapidly, even though the data are well into the
nonlinear regime. Since the above analysis says that
clustering should then evolve more slowly, this has led
some (e.g. Efstathiou et al. 1991) to suggest that the
galaxies seen in the faint surveys are a different population,
with much weaker intrinsic clustering.

\ig{11}{1}
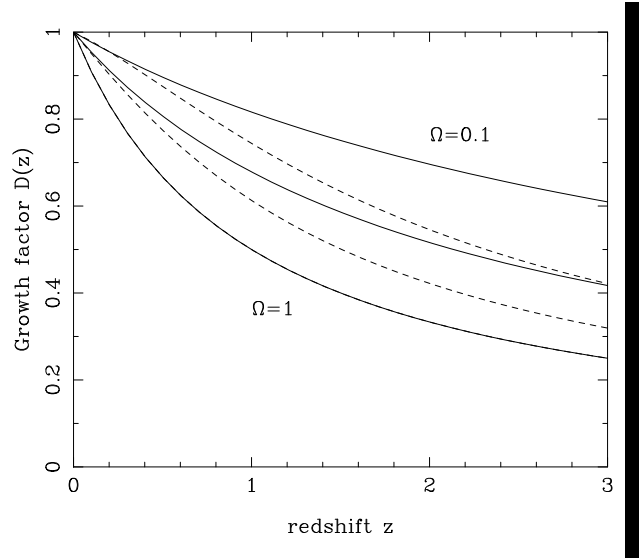
{The linear growth in density perturbations, scaled to
unity at the present, for $\Omega=1$, 0.3 \& 0.1. Open
models are shown as solid lines, flat models are dashed.}

However, there is a serious loophole in the stable-clustering
argument, since we have seen that the observations of clustering
mainly do not reach the stable-clustering regime. In the intermediate
quasilinear transition to linear behaviour, PD96 have
shown that the steep form of $f_{\ss NL}$ produces evolution 
of clustering that is much more rapid:
$$
\xi(r,z) = [r/r_0]^{-\gamma}\, [D(z)]^{(6-2\gamma)(1+\alpha)/3},
$$
where $\alpha\simeq 3.5$ -- 4.5 is the transition slope in 
$f_{\ss NL}$ and $D(z)$ is the linear-theory density growth
rate ($D(z)=(1+z)^{-1}$ if $\Omega=1$). For $\alpha=4$, $\gamma=1.7$,
this gives $\xi\propto D^{4.3}$, which is about twice 
the observed evolution if $\Omega=1$.
Ironically, we still conclude that the observed clustering evolution is
inconsistent with nonlinear evolution in an $\Omega=1$
universe -- but now because the real evolution is too slow, not too fast. 
Agreement with the data requires a linear growth rate which is
about half as rapid as the $\Omega=1$ $D(z)\propto (1+z)^{-1}$.
Figure 11 shows the linear-theory growth factor plotted against
scale factor for various values of $\Omega$ for open and flat
models (see e.g. Carroll et al. 1991). 
This suggests that the observed rate of evolution can be supplied naturally by
$\Omega\simeq 0.2$ -- 0.3 [open] or 0.1 -- 0.2 [flat], without having to postulate that
the faint galaxies are a separate population.

We can do better than this general discussion by looking
at the detailed data on the evolution of clustering from the
Canada-France Redshift Survey (Le F\`evre et al. 1996).
From a sample of 591 galaxies selected to $I\ls 22.5$, the
CFRS team are able to deduce the spatial clustering in several
redshift bands out to $z=1$.
They avoid redshift-space effects by working with the
projected correlation function discussed in the context of
IRAS galaxies. Le F\`evre et al. call this quantity $w(r_p)$, but
we shall use the symbol $\Xi(r)$ as before.
The only important difference is that Le F\`evre et al. quote
their results as a function of proper separation, whereas 
comoving units are preferred here. It will also be more convenient to
use the dimensionless function $\Xi(r)/r$, which is then directly
related to overdensity and so does not change as the length units
are scaled from proper to comoving.

The other technicality to worry about is that the deduced
values of $\Xi(r)/r$ depend on the cosmological model used.
The CFRS results are quoted for $\Omega=1$;
inspection of the integral relating $\Xi$ to
$\xi$ says that the projected results must scale in the
following way for other cosmologies:
$$
{\Xi_1(r_1)\over r_1}=
{d\chi_1\over d\chi_\Omega}\;
{D_\Omega\over D_1}\;
{\Xi_\Omega(r_\Omega)\over r_\Omega},
$$
where the increment of comoving distance is
$$
R_0 d\chi={[c/H_0]\; dz\over \sqrt{ \Omega_v + \Omega_m(1+z)^3
+ (1-\Omega_m-\Omega_v)(1+z)^2}},
$$
$D=R_0 S_k(\chi)$ is comoving angular-diameter distance,
and $R_0=(c/H_0)|1-\Omega_m-\Omega_v|^{-1/2}$.
At a given $r$, $\Xi/r$ also depends on cosmology
because $r_1/r_\Omega=D_1/D_\Omega$. Overall, this shift
and the multiplicative factors cause a larger $\Xi/r$
to be inferred for low $\Omega$. Assuming a slope of
1.65, the total shift of power at $z=1$ is a factor
1.4 ($\Omega=0.3$ open) or 1.9 ($\Omega=0.3$ flat);
these models require an evolutionary $\epsilon$
which is smaller by 0.5 and 0.9 respectively.
There is thus something of a cosmic conspiracy: the low-density
models in which dynamics yields less rapid clustering
evolution are also those in which  geometry causes
less rapid evolution to be inferred from a given set of data.

\beginfigure{12}
\epsfxsize=7.0cm
\epsfbox[81 15 434 790]{pfig12.ps} 
\caption{%
{\bf Figure 12.}
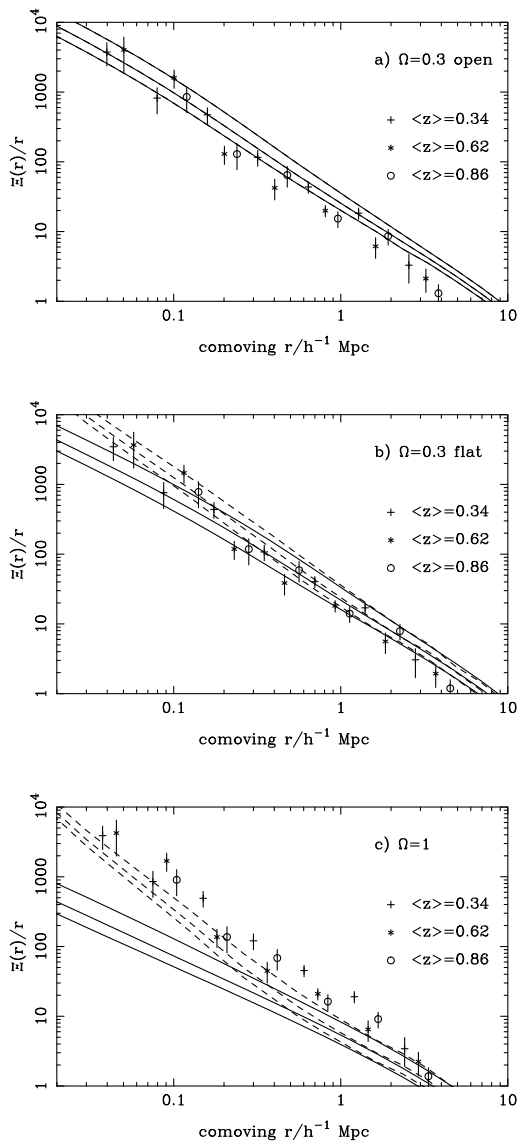
The CFRS projected clustering data (calculated
as a function of comoving separation for three cosmologies) in three
redshift ranges, compared to model predictions.
The solid lines show the evolution predicted for the
double power-law fits to the linear power spectra shown in Fig. 8.
The dashed lines show the effect of adding an additional
small-scale component to the linear spectrum, such as is needed
to account for the present-day clustering spectrum on the smallest scales.
In the case of the $\Omega=1$ model, 
$b=1.6$ at the present was assumed in order to obtain the mass spectrum.
However, unless there is an additional small-scale component,
the implied small-scale bias at high redshifts is very much
greater than this,  on scales where the spectrum is
only mildly nonlinear.}

\endfigure

\ssec{Comparison of models and CFRS data}

The results of a comparison between the CFRS data and various
models are shown in Fig. 12. This shows that the projected
correlations are very close to a single power law on all
nonlinear scales, and that the amplitude changes by only a factor
$\simeq 2$ between $z=0.34$ and $z=0.86$. The solid lines in Fig. 12
are the predictions of the double power-law 
spectra deduced from the local data, and they
show that it is not easy to satisfy the twin constraints of
the right shape and the right rate of evolution. All models
tend to flatten on small scales as clustering reaches the
virialized regime. For $\Omega=0.3$ (open), this tendency is
relatively minor, but for flat models it is a stronger effect,
and produces a huge discrepancy in the case of $\Omega=1$.
If we are to live in an $\Omega=1$ universe, then the
two powers of ten discrepancy in small-scale clustering at
high redshift revealed by Fig. 12c must be bridged by bias.
This is implausible, not just because of the large
amplitude of the bias, but because of the scale on which
it occurs. If we ask what power spectrum is required to
produce the observed power-law clustering, and compare it with
the $\Omega=1$ nonlinear prediction at $z=0.86$, there is a
discrepancy of a factor $\sim 100$ at the point where the
prediction has $\Delta^2\sim 1$. This conflicts with the
general idea that local bias schemes would yield a
near-constant bias where the mass fluctuations are in the
linear regime. If bias of this magnitude were a reality,
it would mean that the galaxy distribution at high redshifts is
dominated by non-local effects which bear little relation to 
the mass distribution. The known existence of massive
high-redshift clusters (Luppino \& Gioia 1995) is probably 
sufficient to rule out this idea.

Alternatively, we may take seriously the small-scale `kick-up' in the
linear spectrum that is required if the idea of strongly scale-dependent
bias is rejected. 
The dashed lines in Fig. 12 show the results of adding in
this feature, which changes the picture so that
the variation between different cosmologies is now much reduced.
The hypothetical additional component dominates
the clustering in the $\Omega=1$ case, and greatly increases the
small-scale clustering in the $\Lambda$-dominated universe.
We now see the standard
stable-clustering evolution on most scales, with something
more rapid on the largest scales. 
The stable-clustering rate in the
$\Lambda$-dominated case is in reasonable accord with the data
as calculated for this model, as expected from the above discussion.
The same would also be true of the $\Omega=1$ model
if the same degree of bias ($b=1.6$) applied at $z=1$ as at $z=0$.
However, the model that comes closest to matching the overall 
shape and normalization of high-redshift
clustering is the open universe. 
The main difficulty for this
model is that it over-predicts the clustering at $z=0.34$, as
does the flat model. Le F\`evre et al. point out that the
galaxies in this bin are of low luminosity only, so this
may not be a problem. The highest-redshift bin contains the
galaxies most nearly comparable to the local samples, and this
is well matched in amplitude if $\Omega\simeq 0.3$.
The preferred range for $\Omega$ depends on how seriously
the CFRS errors are taken, but models outside the range
$\Omega = 0.2$ -- 0.5 give a poor fit
if we stick to the assumption that
there is little bias.

Given the realistic uncertainties in the clustering
data, there is not much to choose between
these latter alternatives. Le F\`evre et al. find a difference in
clustering amplitude of a factor 1.8 between their red and blue
sub-samples at intermediate redshifts, and the difference
between models is of this order. 
One way of progressing from this point will be via
data at higher redshifts, where the evolution of
the different alternatives continues to diverge.
Fern\'andez-Soto et al. (1996) have argued from clustering of
quasar absorption-lines that the scale-length for the correlation
function at $z=2.6$ is $r_0\simeq 0.4 \mpcoh$ (for $\Omega=1$).
The models with additional small-scale features in the
linear spectrum would predict a much larger number, since
evolution would have to proceed at the stable-clustering rate.
If we could be convinced that the absorbers are not strongly
antibiased, this result would therefore argue in favour of open models.

To sum up the conclusions of this Section, 
open universes with a linear spectrum in the form of
a simple break between two power laws predict the
correct form and rate of evolution of galaxy clustering
up to $z=1$. To obtain competitive results in
$\Omega=1$ or flat universes, we must
believe that the linear power spectrum contains an
extra feature, which comes to dominate for $k\gs 1 \hompc$.

\sec{CMB ANISOTROPIES}

A consistent model must match the normalization of the mass 
fluctuations on large scales inferred from fluctuations in
the Cosmic Microwave Background. In making this comparison,
it is important to be clear that the CMB fluctuations
depend only on the very large-scale $P\propto k^n$ portion
of the spectrum. Predictions of smaller-scale
fluctuations such as the amplitude $\sigma_8$ then
need additional information in the form of
the transfer function.
Rather than quoting the $\sigma_8$ implied by the CMB, it
is therefore clearer to give the large-scale normalization
separately.

The large-scale normalization from
the 2-year COBE data in the context of CDM-like models
is discussed by Bunn, Scott \& White (1995);
White \& Bunn (1995); and Stompor, G\'orski \& Banday (1995).
The final 4-year COBE data favour slightly lower results
(Bennett et al. 1996), and
we scale to these in what follows.
For scale-invariant spectra and $\Omega=1$, the best 
normalization of the primordial spectrum is
$$
\Delta^2(k)=(k/0.0737\, h\, {\rm Mpc}^{-1})^4,
$$
equivalent to $Q_{\rm rms}=18.0\, \mu{\rm K}$, or $\epsilon=3.07\times 10^{-5}$
in the notation of Peacock (1991), with an rms error in density fluctuation
of 8\%.
For low-density models, a naive analysis as in PD94 suggests that the
power spectrum should depend on $\Omega$ and the growth factor $g$ as
$P\propto g^2/\Omega^2$. Because of time dependence of gravitational
potential (integrated Sachs-Wolfe effect) and spatial curvature, this
expression is not exact, although it captures the main effect. From the
data of White \& Bunn (1995), a better approximation is
$$
\Delta^2(k)\propto {g^2\over \Omega^2}\; g^{0.7}.
$$
This applies for low-$\Omega$ models both with and without
vacuum energy, with a maximum error of 2\% in density fluctuation
provided $\Omega\ge 0.2$ (and gives the same $\sigma_8$ values as
G\'orski et al. (1995), when the appropriate $\Gamma^*$ corrections
are made, to within 3\%).
Since rough power-law approximations for $g$ are
$g\simeq \Omega^{0.65}$ and $\Omega^{0.23}$ for open and flat
models respectively, we see that the implied density fluctuation
amplitude scales approximately as $\Omega^{-0.12}$ and $\Omega^{-0.69}$
for these two cases. The dependence is very weak for open models,
but vacuum energy implies very much larger fluctuations.
Note that these conclusions can be reached without being specific
about the generation of anisotropies in open models. The existence
of a curvature scale destroys the possibility of a unique
scale-invariant spectrum, but such a concept does have a meaning
on smaller scales. Since the main COBE signal comes from multipoles
with $\ell \gs 10$, the ambiguity in the predicted power from
very low $\ell$ is unimportant unless $\Omega \ls 0.1$.

To avoid large uncertainties caused by different transfer functions,
it is interesting to compare this COBE normalization with
the observed power on the largest reliable scales:
$$
\Delta^2_{\rm opt}(k=0.02 \hompc) \simeq 0.005.
$$
For this scale, the COBE scale-invariant prediction
is 0.0054. Reducing $\Omega$
boosts this by a factor of 1.3 (open $\Omega=0.3$) or
5.3 (flat $\Omega=0.3$). The open number is well
consistent with the observations, given the uncertainty
in the transfer function at this point;
the $\Omega=1$ number is about three times too high (because the optical
power should be roughly $1.6^2$ times that of the mass in the
$\Omega=1$ case); the flat number is too high by at least a factor 6. The last
two cases would therefore require a tilted spectrum.

According to the simplified analysis of PD94,
COBE determines the spatial power spectrum at an
effective wavenumber of $0.0012\,\Omega^\delta\,h\,{\rm Mpc}^{-1}$,
where $\delta=1$ for open models, 0.4 for flat.
The tilt needed to match the power at $k=0.02\hompc$ is then
$n=0.64$ for $\Omega=1$, or $n=0.48$ for the flat $\Omega=0.3$ case.
If gravity waves are included with the usual inflationary
coupling between wave amplitude and tilt, the effect is approximately
$$
\Delta^2 \propto [1+6(1-n)]^{-1}
$$
Allowing for gravity waves therefore changes the required tilt to
$n=0.85$ ($\Omega=1$) or $n=0.75$ ($\Omega=0.3$ flat), but the conclusion
remains that low-density flat models need an extremely large
degree of tilt in order to be viable.
An open model therefore gives the most straightforward match between
COBE and large-scale structure. It is interesting to note that
Ratra et al. (1996) and G\'orski et al. (1996)
also claim that the degree-scale anisotropy
data are best fitted by an open model with $\Omega=0.3$ -- 0.4.

\sec{DISCUSSION} 

This paper has argued that the state of observations in
galaxy clustering is now one of very high precision,
both in the local power spectrum and in its evolution.
Comparing this body of data with the predictions of
nonlinear gravitational instability has yielded a
number of general conclusions:

\item{(1)}
There is some evidence that more power is measured
in redshift space than in real space, but not by
as large a factor as expected for $\beta=1$.
A value $\beta\simeq 0.5$ is more consistent with the value of $\sigma_8$
inferred from the cluster abundance.

\item{(2)}
The shape of the power spectrum is not consistent
with nonlinear evolution of any CDM-like model.
The linear spectrum must have a sharp break around
$k\simeq 0.1\hompc$.

\item{(3)}
Only low-density open models give a small-scale
spectrum which is a power law. For other models, either
a feature in the linear spectrum or an abrupt increase
in small-scale bias is needed.

\item{(4)}
Open models with $\Omega\simeq 0.3$ also naturally give a spectrum which
evolves at the observed rate while maintaining the
same power-law shape. Such models are furthermore the
only alternatives that are consistent with CMB
anisotropies without requiring a large tilt.

These conclusions have been reached without invoking a specific
mechanism for how galaxies trace mass. For low-density universes,
we have been content to use $M/L$ arguments to infer that the
empirical degree of bias today is probably small, whereas
substantial bias must exist if $\Omega=1$. A more complete
picture would have to be able to calculate the amount of
bias and how it changes with redshift, but this is
not well constrained at present. For example, Efstathiou (1995) 
showed how faint-galaxy clustering of the required amplitude
could exist at $z=1$ in an $\Omega=1$ CDM universe
by explicitly following the formation of
low-mass haloes of dark matter in a numerical simulation, and
identifying these with galaxies at that time.
The problem with this argument is that different choices of
halo mass (and halo environment) produce huge variations in
the clustering amplitude, so this route does not yet
give a robust prediction for the clustering of faint galaxies.
This ability to obtain very different clustering
properties from different plausible ideas for how and
where galaxies form warns us to take great care in
interpreting clustering data;
even so, it is undeniably striking how closely
an open universe in which galaxies nearly trace mass accounts for
a variety of aspects of galaxy clustering and its evolution.

\section*{ACKNOWLEDGEMENTS}

Thanks are due to Carlton Baugh and Helen Tadros for communicating APM
power-spectrum data, and to Will Saunders for
illuminating conversations on the relation between
optical and IRAS clustering.

\section*{REFERENCES}

\beginrefs

\ref Albrecht A., Stebbins A., 1992, Phys. Rev. Lett., 69, 2615
\ref Ballinger W.E., Peacock J.A., Heavens A.F., 1996, \mn, in press
\ref Baugh C.M., Efstathiou G., 1993, \mn, 265, 145
\ref Baugh C.M., Efstathiou G., 1994, \mn, 267, 323
\ref Bennett C.L. et al., 1996, \apj, 464, L1
\ref Branchini E., Guzzo L., Valdarnini R., 1994, \apj, 424, L5 
\ref Bunn E.F., Scott D., White M., 1995, \apj, 441, 9
\ref Carroll S.M., Press W.H., Turner E.L., 1992, \annrev, 30, 499
\ref Cole S., Fisher K.B., Weinberg D.H., 1995, \mn, 275, 515
\ref Coles P., 1993, \mn, 262, 1065
\ref Couch W.J., Jurcevic J.S., Boyle B.J., 1993. \mn, 260, 241
\ref Davis M., Peebles  P.J.E., 1983 \apj, 267, 465
\ref Dekel, A., 1994, \annrev, 32, 371
\ref de Lapparent V., Geller M.J., Huchra J.P., 1986, \apj, 302, L1
\ref Efstathiou G., Bond J.R., 1986, \mn, 218, 103
\ref Efstathiou G., Bernstein G., Katz N., Tyson T., Guhathakurta P., 1991, \apj, 380, 47
\ref Efstathiou G., Bond J.R., White S.D.M., 1992, \mn, 258, 1P
\ref Efstathiou G., Sutherland W.J., Maddox S.J., 1990, \nat, 348, 705
\ref Efstathiou G., 1995, \mn, 272, L25
\ref Feldman H.A., Kaiser N.,  Peacock J.A., 1994, \apj, 426, 23
\ref Fern\'andez-Soto A., Lanzetta K.M., Barcons X., Carswell R.F., Webb J.K., Yahil, A., 1995, \apj, 460, L85
\ref Fisher K.B., Davis M., Strauss M.A., Yahil A.,  Huchra J.P., 1993, \apj, 402, 42
\ref Fisher K.B., Davis M., Strauss M.A., Yahil A.,  Huchra J.P., 1994, \mn, 267, 927
\ref Fisher K.B., Nusser A., 1996, \mn, 279, L1
\ref G\'orski K.M., Ratra B., Sugiyama N., Banday A.J., 1995, \apj, 444, L65
\ref G\'orski K.M., Ratra B., Stompor R., Sugiyama N., Banday A.J., 1996, astro-ph/9608054
\ref Hamilton A.J.S., Kumar P., Lu E.,  Matthews A., 1991, \apj, 374, L1 
\ref Holtzman J.A., 1989, \apjs, 71, 1
\ref Jain B., Mo H.J., White S.D.M., 1995, \mn, 276, L25 
\ref Kaiser N., 1987, \mn, 227, 1
\ref Klypin A., Holtzman J., Primak J., Reg\H os E., 1993, \apj, 416, 1
\ref Klypin A.,  Primak J., Holtzman J., 1996, \apj, 466, 13
\ref Le F\`evre O., et al., 1996. \apj, 461, 534
\ref Loveday J., Efstathiou G., Peterson B.A., Maddox S.J., 1992, \apj, 400, L43
\ref Loveday J., Maddox S.J., Efstathiou G., Peterson B.A., 1995, ApJ, 442, 457
\ref Loveday J., Efstathiou G., Maddox S.J., Peterson B.A., 1996, ApJ, in press
\ref Luppino, G., Gioia I., 1995, \apj, 445, L77
\ref Maddox S.J., Efstathiou G., Sutherland W.J., Loveday J., 1990, \mn, 242, 43P
\ref Maddox S.J., Efstathiou G., Sutherland W.J., 1996, \mn, in press
\ref Mann R.G., Peacock J.A., Heavens A.F., 1996, \mn, in preparation.
\ref Ma C.P., Bertschinger E., 1994, \apj, 434, L5
\ref Mo H.J., Jing Y.P., B\"orner G., 1993, \mn, 264, 825
\ref Mo H.J., Miralda-Escud\'e J., 1994, \apj, 430, L25
\ref Mo H.J., Fukugita M., 1996, \apj, 467, L9
\ref Neuschaefer L.W., Windhorst R.A., Dressler A., 1991, \apj, 382, 32
\ref Peacock J.A., 1991, \mn,  253, 1P
\ref Peacock J.A., Dodds S.J., 1994, \mn, 267, 1020 (PD94)
\ref Peacock J.A., Dodds S.J., 1996, \mn, 280, L19 (PD96)
\ref Peebles P.J.E., 1980, The Large-Scale Structure of the Universe.  Princeton Univ. Press, Princeton, NJ
\ref Pogosyan D.Y., Starobinsky A.A., 1995, \apj, 447, 465
\ref Ratcliffe A., Shanks T., Broadbent A., Parker Q.A., Watson F.G., Oates A.P., Fong R., Collins C.A., 1996, \mn, 281, L47
\ref Ratra B., Banday A.J., G\'orski K.M., Sugiyama N., 1996, preprint PUPT-1558
\ref Roche N., Shanks T., Metcalfe N., Fong R., 1993, \mn, 263, 360
\ref Saunders W., Frenk C., Rowan-Robinson M., Efstathiou G.,     Lawrence A., Kaiser N., Ellis R., Crawford J., Xia X.-Y., Parry I., 1991, \nat, 349, 32
\ref Saunders W., Rowan-Robinson M., Lawrence A., 1992, \mn, 258, 134
\ref Stompor R., G\'orski K.M., Banday A.R., 1995, \mn, 277, 1225
\ref Strauss M.A., Davis M., Yahil Y., Huchra J.P., 1992, \apj, 385, 421
\ref Sugiyama N., 1995, ApJ Suppl, 100, 281
\ref Tadros H., Efstathiou G., 1995, \mn, 276, L45
\ref Tadros H., Efstathiou G., 1996, \mn, in press
\ref Taylor A.N., Rowan-Robinson M., 1992, \nat, 359, 396
\ref Vogeley M.S., Park C., Geller M.,  Huchra J.P., 1992, \apj, 391, L5
\ref White M., Bunn E.F., 1995, \apj, 450, 477
\ref White S.D.M., Efstathiou G., Frenk C.S., 1993, \mn, 262, 1023

\endrefs
\bye